\documentclass[fleqn,usenatbib,onecolumn]{mnras}

\usepackage{newtxtext,newtxmath}
\usepackage{xcolor}
\usepackage{float}
\usepackage{soul}

\usepackage[T1]{fontenc}
\usepackage{ae,aecompl}

\usepackage{graphicx}	
\usepackage{amsmath}	
\usepackage{amssymb}	
\usepackage{tablefootnote}
\definecolor{officegreen}{rgb}{0.0, 0.5, 0.0}

\title[Cut-off periods in a solar stratified atmosphere]{Solar slow magneto-acoustic-gravity waves: an erratum correction and a revisited scenario}

\author[Zurbriggen et al.]{E. Zurbriggen$^{1}$\thanks{EZ-mail: \sf{ernesto.zurbriggen@craam.mackenzie.br}}, M. V. Sieyra$^{2}$\thanks{MVS-mail: \sf{vsieyra@unc.edu.ar}. Formerly at Instituto de Astronom\'ia Te\'orica y Experimental (IATE--CONICET), Universidad Nacional de C\'ordoba (UNC), C\'ordoba, Argentina.},  A. Costa$^{3}$\thanks{AC-mail: (CA)  \sf{acosta@oac.unc.edu.ar}}, A. Esquivel$^{3,4}$, G. Stenborg$^{5}$ \\
$^{1}$ Universidade Presbiteriana Mackenzie, Escola de Engenharia, Centro de R\'adio Astronomia e Astrof\'isica Mackenzie (CRAAM), SP, \\ $~$ S\~ao Paulo, Brasil.\\
$^{2}$ Centro de Estudios para el Desarollo Sustentable (CEDS), Universidad Tecnol\'ogica Nacional (UTN)\,--\,Facultad Regional Mendoza. \\
$~$ Consejo Nacional de Investigaciones Cient\'ificas y T\'ecnicas (CONICET). Mendoza, Argentina. \\
$^{3}$ Consejo Nacional de Investigaciones Cient\'ificas y T\'ecnicas (CONICET), Instituto de Astronom\'ia Te\'orica y Experimental (IATE), \\ $~$ C\'ordoba, Argentina. \\
$^{4}$ Instituto de Ciencias Nucleares, Universidad Nacional Aut\'onoma de M\'exico, A. Postal. 70-543, 04510, Ciudad de M\'exico, M\'exico. \\
$^{5}$ Space Science Division, U.S. Naval Research Laboratory, Washington, DC 20375, USA.}

\date{}

\pubyear{2020}

\usepackage{etoolbox}
\makeatletter
\patchcmd\@combinedblfloats{\box\@outputbox}{\unvbox\@outputbox}{}{%
   \errmessage{\noexpand\@combinedblfloats could not be patched}%
}%

\defcitealias{costa2018}{C18}

\begin{document}
\label{firstpage}
\pagerange{\pageref{firstpage}--\pageref{lastpage}}
\maketitle

\begin{abstract}
Slow waves are commonly observed on the entire solar atmosphere. Assuming a thin flux tube approximation, the cut-off periods of slow-mode magneto-acoustic-gravity waves that travel from the photosphere to the corona were obtained in \cite{costa2018}. 
In that paper, however, a typo in the specific heat coefficient at constant pressure $c_{\mathrm{p}}$ value led to an inconsistency in the cut-off calculation, which is only significant at the transition region.
Due to the abrupt temperature change in the region, a change of the mean atomic weight (by a factor of approximately two) also occurs, but is often overlooked in analytical models for simplicity purposes.
In this paper, we revisit the calculation of the 
cut-off periods of magneto-acoustic-gravity waves in \cite{costa2018} by considering an atmosphere in hydrostatic equilibrium with a temperature profile, with the inclusion of  the variation of the mean atomic weight and the correction of the inconsistency aforementioned. In addition, we  show that the cut-off periods obtained analytically are consistent with the corresponding periods measured in observations of a particular active region.

\end{abstract}

\begin{keywords}
magnetohydrodynamics -- waves -- analytical
\end{keywords}


\section{Introduction}

Magnetohydrodynamic oscillations and waves are ubiquitous in the solar atmosphere, from the photosphere up to the corona, and they show a wide range of periods. Changes in plasma parameters, e.g. density, temperature and magnetic field, influence the propagation of such waves affecting their physical properties as intensity, propagation speed and mode conversion.

In particular, the presence of slow magneto-acoustic-gravity waves (MAG waves, hereafter) guided by the magnetic field is supported by increasingly observational evidence in different atmospheric structures, e.g. photospheric flux tubes \citep{1997LNP...489...75R}, sunspots \citep{2015LRSP...12....6K,madsen2015,freij2014,jess2013}, coronal loops \citep{2003A&A...404L...1K} and coronal plumes \citep{2006RSPTA.364..473N}. 
These oscillations are also used to estimate the formation heights of different emission spectral lines.  
As a result of the analysis of 3 minute oscillations detected in observations from the Atmospheric Imaging Assembly instrument \citep[AIA,][]{lemen2012} on-board the Solar Dynamic Observatory \citep[SDO,][]{pesnell2012}, \citet{2015ARep...59..959D} found that the formation heights of the corresponding spectral lines are consistent with models of the sunspots umbra involving strong temperature gradients of the type used in this paper \citep[e.g.,][]{2009ApJ...707..482F}.
Meanwhile the lower atmospheric magnetic structure is frequently believed to be mostly formed by small magnetic flux tubes of circular cross-section emerging from the photosphere and expanding upwardly in the corona \citep{solanki2006}. The photosphere and chromosphere are known to be dominated by acoustic-gravity waves with observed short periods of $\sim$[3--5]$~$min. Whereas in the lower corona a wider range of periods are present (up to about 80$~$min --e.g., \citealt{2002SoPh..209..265S}--), short periods ([3--5] min) have also been detected in certain coronal structures, e.g., coronal loops and intense magnetic flux tubes \citep{reznikova2012,jess2012,2010NewA...15....8S}. A rich range of periods are found in sunspots, 
the periods becoming larger as the distance from the sunspot umbra increases; a likely result of the influence of the magnetic field 
(\citealt{madsen2015}; \citealt{yuan2014}; \citealt{jess2013}; \citealt{2006RSPTA.364..313B}).

The cut-off frequency of an oscillating system is the frequency at which the wave becomes evanescent, incapable of transporting energy. \citet{1932hydr.book.....L} showed that this cut-off frequency may be the result of the natural response of the solar atmosphere to disturbances and interpreted the observations as the normal mode oscillations of the system. The cut-off frequency for MAG waves in an isothermal medium with a uniform magnetic field have been studied by several authors in the past, e.g., \citet{2006roberts,zhugzhda1984,thomas1982}.

The thin flux tube model proposed by \citet{1978SoPh...56....5R} has shown to be appropriate to study propagation of long-wavelength slow MAG waves in the solar atmosphere. By means of this approximation it is possible to obtain the wave equation and the cut-off frequency for slow modes in a vertical magnetic flux tube. More recently, \citet{afanasyev2015} calculated coronal cut-off frequencies considering a divergent magnetic flux tube in an isothermal corona. Following this approach, \citet[][hereafter \citetalias{costa2018}]{costa2018} studied  a flux tube expanding from the photosphere to the corona that included a varying magnetic field, a non-isothermal temperature and a gravitationally stratified atmosphere. As \citet{1932hydr.book.....L}, we interpreted the obtained monochromatic oscillations as the response to perturbations at the natural frequency of the solar atmosphere and suggested that the coincidence between atmospheric and heliospheric frequencies (e.g., p-modes) could be the result of an evolving process rather than the consequence of a force applied at the base of the photosphere. As \citet{afanasyev2015}, \citetalias{costa2018} explained the observed abundance of periods between $\sim$[15--80]$~$min in the corona, and  also explained the short periods of $\sim$[3--5]$~$min detected in the phostosphere.

The cut-off calculation of \citetalias{costa2018} has, however, an inconsistency due to a typo in the value of the specific heat coefficient $c_{\mathrm{p}}$ (at constant pressure), which is mainly significant at the transition region and may be appreciated in the maximum period values of their figure 4. In the present work we fix the inconsistency, revise and improve the scenario allowing the mean atomic weight to vary, and compare our analytical results with measurements of photospheric and coronal periods in an active region. The mean atomic weight initially represents a weakly ionised photosphere, then varies with height  through the transition region, and ends representing a fully ionised corona. One important consequence of taking into account a varying mean atomic weight is that the variation of the equilibrium atmospheric density profile is accurately considered.
Our model assumes a (steady state) temperature stratification with height, and does not include the detailed balance between heating and cooling processes. In other words, the model assumes that the heating/cooling timescales are much shorter than the time variation of the other physical quantities \citep[see discussion in][]{2019PhPl...26h2113Z}. We must note that we are also restricting our modelling to ideal (neither viscous nor resistive) processes alone. This could potentially affect the propagation of waves in the plasma, but that is beyond the scope of this work. For instance, we have considered the variations of the mean atomic weight with height due to the change in ionisation fraction, but we neglect the effect of ambipolar drift which would act as a source of viscous damping of the waves  \citep{2017PPCF...59a4038K}.

Finally, an explicit expression of the cut-off frequency is provided as general as possible in the framework of the approximations assumed, valid for an arbitrary equilibrium atmosphere considering a thin flux tube model.

\section{Wave equation and cut-off period}

Here we study the cut-off period of slow MAG waves propagating outwards from the photosphere to the lower corona within a flux tube. Considering the thin flux tube approximation we derive a Klein-Gordon equation that explicitly contains the cut-off frequency. This approximation describes the plasma dynamics in an intense, straight, untwisted and non-rotating flux tube.

Given the wave characterisation we are addressing, we neglect dispersion effects caused by the finite radius of the flux tube. In other words,  the flux tube is assumed narrow, in the sense that the radial scale variation $R$ is negligible with respect to the vertical scale-length $L$ (i.e. $R/L\ll 1$). For this, a first non-zeroth order Maclaurin radial expansion for the MHD equations is valid (see the discussion of \citealt{zhugzhda1996}). The wave-guiding  condition of these plasma structures imply that the slow MAG modes manifest themselves as  tube modes. The plasma contained in the tube is compressible, without a steady flow, in hydrostatic equilibrium with a stratification in the vertical direction that includes a temperature profile and a varying mean atomic weight.

The procedure is as follows: a system of cylindrical coordinates $(r,\varphi,z)$ for the radial, azimuthal and vertical (outward from the sun surface) directions is used. Within the general set of ideal MHD equations with a gravitational field in the $z$-axis  we set the following assumptions:
(i)  the flux tube is narrow;
(ii)  not twisted, therefore it has azimuthal symmetry;
(iii) the magnetic field outside the tube is negligible compared with that in the flux tube (magnetically intense).
Also  the perturbations in the external medium are assumed negligible only supporting evanescent disturbances \citep{1979SoPh...64...77R}.

With this considerations, the classic thin flux tube approximation provides a set of $1.5$D MHD equations, which expressed in the variables $(\rho,p,u,B)$, is given by:
\begin{eqnarray} \label{eq:tf}
\begin{aligned}
\frac{\partial \rho}{\partial t} + 2 \rho v + \frac{\partial}{\partial z} (\rho u) & =  0, \\
\rho \left( \frac{\partial u}{\partial t} + u \frac{\partial u}{\partial z} \right) & = - \frac{\partial p}{\partial z} - \rho g, \\
\frac{\partial B}{\partial t} + 2vB + u \frac{\partial B}{\partial z} & = 0, \\
\frac{\partial p}{\partial t} + u\frac{\partial p}{\partial z} & = \frac{\gamma p}{\rho}\left(\frac{\partial \rho}{\partial t} + u\frac{\partial \rho}{\partial z}\right), \\
\end{aligned}
\end{eqnarray}
where $\rho$ is the mass density, $p$  the thermal pressure, $u$ the longitudinal (vertical) component of the plasma velocity, $v$ the radial component of the velocity, $g$ the solar surface gravity acceleration (assumed constant, in the $z$ direction), $B$ is the longitudinal component of the magnetic field and $\gamma=5/3$.

In this study a general $z$-dependent ideal equation of state is assumed,
\begin{eqnarray} \label{eq:gideal}
p(z) = R_{\mathrm{g}}\frac{\rho(z) T(z)}{\mu(z)},
\label{eq:eos}
\end{eqnarray}
with $T$ being the temperature, $\mu(z)$ the mean atomic weight and $R_{\mathrm{g}}$ the gas constant.

Let us consider linear perturbations of Eqs. (\ref{eq:tf}) around an equilibrium state as:
\begin{eqnarray}  \label{eq:lineal}
\begin{aligned}
& \rho(z,t) = \rho_0(z) + \rho_1(z,t),  \\
& p(z,t) = p_0(z) + p_1(z,t),   \\
& B_z(z,t) = B_0(z) + B_1(z,t), \\
& v(z,t) = v_1(z,t), \\
& u(z,t) = u_1(z,t),
\end{aligned}
\end{eqnarray}
where the subscript $0$ stands for the equilibrium state, and $1$ for the equilibrium departures. Note that $v_0(z)$ and $u_0(z)$ are assumed to be zero. To simplify the notation we will use simply $u_1\equiv u$ and $v_1 \equiv v$.

Replacing Eqs.(\ref{eq:lineal}) into Eqs.(\ref{eq:tf}) and keeping only first-order terms, after some rearrangement the set of equations is reduced to:
\begin{eqnarray}
\frac{B_0}{4\pi}\frac{\partial B_1}{\partial t} + c_0^2\frac{\partial \rho _1}{\partial t} - up_0' + c_0^2 u \rho_0' = 0, & \label{eq:ln1} \\
\rho_0 \frac{\partial u}{\partial t} + \rho_1 g - \frac{B_0}{4 \pi} \frac{\partial B_1}{\partial z} - \frac{B_1}{4 \pi} B_0' = 0, & \label{eq:ln2}  \\
\rho_0 \frac{\partial B_1}{\partial t} +  \rho_0 B_0' u- B_0 \frac{\partial\rho_1}{\partial t} - B_0\rho_0' u - B_0\rho_0 \frac{\partial u}{\partial z} = 0. &  \label{eq:ln3}
\end{eqnarray}
The prime symbol $(')$ denotes differentiation with respect to height ($\equiv d/d z$).

Finally, a wave equation  for the longitudinal speed $u$ is obtained combining the time derivative of Eq. (\ref{eq:ln2}) with the spatial derivative of Eq. (\ref{eq:ln3}) and grouping the terms with $u$ and $\partial u/\partial t$:
\begin{eqnarray} \label{eq:onda}
\frac{\partial^2 u}{\partial t^2} - c_{\mathrm{T}}^2 \frac{\partial^2 u}{\partial z^2} + K_1(z) \frac{\partial u}{\partial z}  + K_2(z) u = 0.
\end{eqnarray}
where
\begin{eqnarray}
c_0^2 = \frac{\gamma p_0}{\rho_0}, \qquad V_{\mathrm{A}}=\frac{B_0}{\sqrt{4\pi\rho_0}}, \qquad c_{\mathrm{T}} = \frac{c_0 V_{\mathrm{A}}}{\sqrt{c_0^2+V_{\mathrm{A}}^2}},
\end{eqnarray}
are the equilibrium sound speed, Alfv\'en speed and tube speed, respectively. The  $K_1(z)$ and $K_2(z)$ functions are:
\begin{eqnarray}
\begin{aligned}
K_1  = \ &  c_{\mathrm{T}}^2 \frac{V_{\mathrm{A}}^2-c_0^2}{V_{\mathrm{A}}^2+c_0^2}\frac{B_0'}{B_0} + g \gamma \frac{c_{\mathrm{T}}^4}{c_0^4}, \\
K_2  = \ & c_{\mathrm{T}}^2 \frac{B_0''}{B_0} - c_{\mathrm{T}}^2 \frac{V_{\mathrm{A}}^2 - c_0^2}{V_{\mathrm{A}}^2 + c_0^2}\left(\frac{B_0'}{B_0}\right)^2 + g\frac{c_T^4}{c_0^4} \left(3\frac{c_0^2}{V_A^2}+1-\gamma\right)\frac{B_0'}{B_0} \\ & + \frac{c_{\mathrm{T}}^2}{V_{\mathrm{A}}^2} N^2 + \gamma g^2 \frac{c_{\mathrm{T}}^4}{V_{\mathrm{A}}^2 c_0^4}.
\end{aligned}
\end{eqnarray}
In this case the equilibrium Brunt-V\"ais\"al\"a frequency $N$ is:
\begin{eqnarray} \label{eq:bb}
N^2(z) = g \left(\frac{1}{\gamma} \frac{p_0'}{p_0} - \frac{\rho_0'}{\rho_0} \right) \equiv (\gamma -1)\frac{g^2}{c_0^2} + g \frac{\left(c_0^2\right)'}{c_0^2}.
\end{eqnarray}
In summary, Eq. (\ref{eq:onda}) establishes the longitudinal dynamic of slow waves in thin magnetic flux tubes allowing the temperature, density, mean atomic weight and magnetic field to vary from the photosphere to the corona.

In order to get the cut-off period, the wave Eq. (\ref{eq:onda}) is reduced to a Klein-Gordon form (the reader is referred to APPENDIX \ref{s:appendix} for further details). Then, the resulting cut-off frequency squared is given by:
\begin{eqnarray} \label{eq:cutoff}
\omega_u^2(z) = -c_{\mathrm{T}}^2\left( \psi' \right)^2 - c_{\mathrm{T}}^2 \psi'' + K_1 \psi' + K_2,
\end{eqnarray}
where $\psi'= K_1/(2 c_{\mathrm{T}}^2)$, and the cut-off period is $P(z)=2\pi/\omega_u(z)$ (see Eq. \ref{eq:cutoff_gen}). 

For completeness, we analyse some special cases of the wave Eq. (\ref{eq:onda}). If we consider a constant mean atomic weight, $\mu'=0$, we obtain the expression for equation 7 of \citetalias{costa2018}. Although at first glance this assumption does not seem to have an explicit effect on the actual Eq. (\ref{eq:onda}), it implicitly does through the sound speed $c_0$ value, as can be appreciated from the identity
\begin{eqnarray}
\frac{\left(c_0^2\right)'}{c_0^2} = \frac{T_0'}{T_0} - \frac{\mu'}{\mu},
\end{eqnarray}
and through the Alv\'en speed ($V_{\mathrm{A}}$) by the  equilibrium density profile $\rho_0$ (see below). If we  assume an isothermal flux tube ($T_0'=c_0'=0$), we recover the wave equation 4 of \citet{afanasyev2015}. Then, adding the  constant magnetic field constraint for the plasma inside the tube ($B_0'=0$), the expression 3.9 of \cite{2006roberts} is reproduced. Finally, for the limiting case of an infinite magnetic field ($V_{\mathrm{A}}/c_0 \rightarrow \infty$ with $B_0'$ not divergent), the well-known acoustic cut-off frequency for an isothermal-stratified medium is obtained ($\omega_u=\gamma g/(2c_0)$, \citealt{1932hydr.book.....L}). Whereas, the acoustic cut-off frequency for a non-isothermal-stratified medium with varying mean atomic weight $\mu(z)$, is
\begin{eqnarray} \label{eq:acous}
\omega_u^2 = {\left(\frac{\gamma g}{2c_0}\right)}^2 + \gamma g\frac{c_0'}{c_0}.
\end{eqnarray}
%

\section{The Atmospheric Model} \label{sec:atm-model}
Firstly we study slow waves in the vertical range {$[z_0$--$z_\mathrm{f}]=[0$--$10]~$Mm}, $z_0$ being an arbitrary reference level fixed at the base of the photosphere. The resulting equilibrium density considering the thin flux tube Eq. (\ref{eq:tf}) is:
\begin{eqnarray} \label{eq:hseq}
\begin{aligned}
& \rho_0(z) = \rho_0(z_0) \frac{c_0^2(z_0)}{c_0^2(z)} \frac{\mu(z)}{\mu(z_0)}\, \mathrm{e}^{-\int_{z_0}^{z} \frac{d\tilde{z}}{H(\tilde{z})}}, \\
& \frac{1}{H(z)} = \frac{\gamma g}{c_0^2} \equiv - \frac{\rho_0'}{\rho_0} - \frac{\left(c_0^2\right)'}{c_0^2}.
\end{aligned}
\end{eqnarray}
$H(z)$ is the atmosphere scale height. The quantities evaluated at $z_0$ are prescribed values.

As previously done in \citetalias{costa2018}, we consider along the tube an equilibrium magnetic field of the form
\begin{eqnarray}
B_0(z)=B_0(z_0)\,\mathrm{e}^{-z/l},
\end{eqnarray}
which is a valid functional form to model a divergent magnetic flux tube. Here, $l$ is the magnetic field scale height and the value at the base is $B_0(z_0)\equiv B_{00}$. 
The radial component of the magnetic field is \citep[see ][]{2017ApJ...840...20S}
\begin{eqnarray}
B_{r0}(r,z) = \frac{r}{2l}B_0 e^{-z/l}
\end{eqnarray}
(a first order radial term with magnitude $B_{r0} \sim \frac{R}{L}B_0$). Thus, magnetic radial direction variations are negligible with respect to longitudinal ones if $R/L\ll 1$. In practise, this condition is satisfied if $l \geq 4 h_\mathrm{b}$, where $h_\mathrm{b}=H(z_0)$ is the scale height evaluated at the base ($h_\mathrm{b}=0.24~$Mm). The field scale parameter $l$ controls the radial expansion of the tube. The smaller the value of $l$, the more divergent the tube, and consequently the larger the decay with height of the magnetic intensity inside the tube. Below the limit of $l$ given by the assumed thin tube condition  ($l \geq 4 h_\mathrm{b}$), the radial and vertical scale-lengths become of the same order of magnitude. The limiting condition for $l$ was calculated considering  the tube radius $R(z)$ as a function of height and using  the conservation of the magnetic flux (inside the tube). We also assumed a typical tube radius at the base of the photosphere of $R(z_0)\approx 0.1~$Mm.

\subsection{Temperature and Mean Atomic Weight}
The equilibrium temperature is introduced in the wave equation through the profile:
\begin{eqnarray}
T_0(z) = a_1 \tanh\left(\frac{z-a_2}{c}\right)+a_3,
\end{eqnarray}
where the parameters $a_{1,2,3}$ and $c$ allow the adjustment to the atmospheric solar temperature profile proposed by \citet{vernazza1981}. The temperature variation chosen, ranges from $T_0(z_0)=10^4~$K at the base of the photosphere to $T_0(z_\mathrm{f})=1.4$$\times$10$^6~$K at  coronal heights. The beginning of the transition region was set at $z=2.0~$Mm.

Different solar temperature profiles were modelled using the parameters shown in Table \ref{t:parameters}.
The parameter $c$ controls the width of the transition region, where the larger  values  correspond to smoother transition regions. We chose two values of $c$, a `sharp' case with $c=0.25~$Mm and a `smooth' case with $c=0.5~$Mm. They correspond to transition region widths of $\sim$1.5$~$Mm and $\sim$4$~$Mm, respectively. Hereafter, we shall refer to these models as the sharp and the smooth case, respectively.
\begin{table*}
\begin{center}
\caption{Parameters for the temperature and mean atomic weight profiles.}
\label{t:parameters}
\vspace{-0.15cm}
\begin{tabular}{ccccccc}
\hline  
$c~$[Mm]$^i$  &     $a_1~$[K]   & $a_2~$[Mm]$^j$ &     $a_3~$[K]     & $b_1^k$ & $b_2~$[Mm] & $b_3^l$ \\ \hline 
0.25          & 7$\times 10^5$  &     12         & 7.1$\times 10^5$  &   0.34  &   2.3      & 0.95  \\ 
0.5           & 7$\times 10^5$  &     22         & 7.1$\times 10^5$  &   0.34  &   2.6      & 0.95 \\ 
\hline
\multicolumn{7}{l}{$^i$ The parameter that controls the width of the mean atomic weight is set to $d=c/2$.} \\
\multicolumn{7}{l}{$^j ~ a_2=4.0c+2.$} \\
\multicolumn{7}{l}{$^k ~ b_1=\frac{1}{2}\left[\mu(z_\mathrm{f})-\mu(z_0)\right]$} \\
\multicolumn{7}{l}{$^l ~ b_3=\mu(z_0) + b_1$} \\
\end{tabular}
\end{center}
\end{table*}

From each temperature profile we compute the mean atomic weight  considering only hydrogen and helium, assuming that the ionisation state is that of coronal equilibrium. Thus, the ionisation fractions of H and He are only function of temperature, and are set to the value that corresponds to the one at each height. The calculation was made with the {\sc chianti} library and its associated {\sc python} software \citep{1997A&AS..125..149D,2019ApJS..241...22D}, assuming an abundance of 90\% H + 10\% He by number. The resulting mean weight stratification has a similar shape (mirrored) as the temperature, thus we fit a profile
\begin{eqnarray} \label{eq.mu}
\mu_(z) = -b_1 \tanh\left(\frac{z-b_2}{d}\right)+b_3,
\end{eqnarray}
where the values of the coefficients are listed in Table \ref{t:parameters} as well. The value of the mean atomic weight at the base of the (weakly ionised) photosphere is $\mu(z_0)$=1.29, and $\mu(z_\mathrm{f})$=0.62 in the fully ionised corona.
Notice that the parameter that determines the width of the $\mu$ profile is different (roughly one half, $d=c/2$) to that of the temperature. This is because at a temperature of $\sim$ 2$\times$10$^5~$K the (H and He) plasma is already fully ionised. In what follows we will use the width parameter of the temperature ($c$) as reference for the different models, keeping in mind that the mean atomic weight has also a sharp/smooth change associated, albeit with a smaller width.

\begin{figure} 
\begin{center}
\includegraphics[width=0.497\columnwidth]{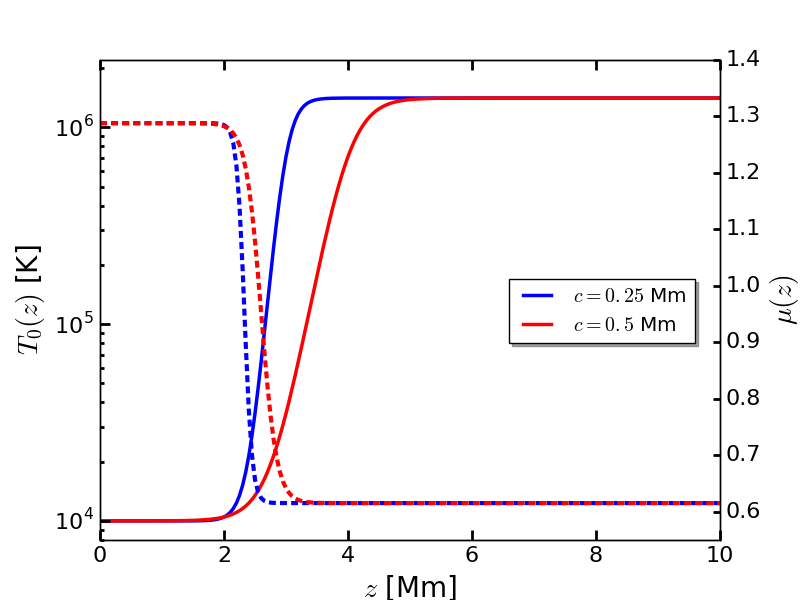}
\includegraphics[width=0.497\columnwidth]{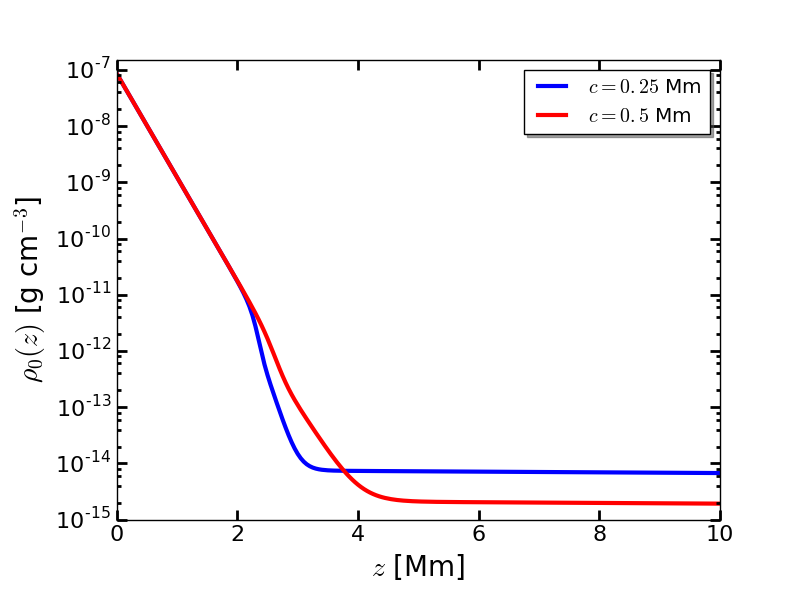}
 \caption{a) Temperature (solid lines, left axis) and mean atomic mass (dashed lines, right axis) profiles as a function of height for two values of $c$ (indicated by the colours, see the legend). The temperature changes from $10^{4}~$K at the photosphere to 1.4$\times$10$^{6}~$K at the corona, while the plasma changes from weakly ionised ($\mu$=1.29) to fully ionised ($\mu$=0.62). b) Density profile for the same $c$ values of the top panel (a), obtained assuming hydrostatic equilibrium (Eq. \ref{eq:hseq}).}
 \label{fig:temp_dens-profile}
\end{center}
\end{figure}

Figure~\ref{fig:temp_dens-profile}a shows the temperature and mean atomic mass profiles for the two values of $c$ mentioned above, with the sharp transition shown in red and the smooth one in blue. The continuous lines (with the left vertical axis) correspond to the temperature stratification, and the dashed lines (with the vertical axis on the right) to the mean atomic mass. The density profile is determined from the  hydrostatic equilibrium {Eq. (\ref{eq:hseq})} and considering the temperature and mean atomic weight stratifications. Figure~\ref{fig:temp_dens-profile}b shows the density as a function of height, for the same values of $c$. Note that the smooth case leads to a smaller coronal density value. At the base it is $\rho_0(z_0)=8.6\times10^{-8}~$g cm$^{-3}$, while high in the corona it is $\rho_0(z_\mathrm{f})=6.7\times10^{-15}~$g cm$^{-3}$ for the sharp case (blue line) and $\rho_0(z_\mathrm{f})=1.9\times 10^{-15}~$g cm$^{-3}$ for the smooth one (red line). Finally, the number density values at the base of the photosphere for hydrogen and helium are $n_{\textrm{H}}=3.6\times10^{16}~$cm$^{-3}$ and $n_{\textrm{He}}=4\times 10^{15}$ cm$^{-3}$, respectively. 
%

\section{Results and discussion}
\subsection{Analytical calculation}
\begin{figure}
\begin{center}
\includegraphics[width=0.55\columnwidth]{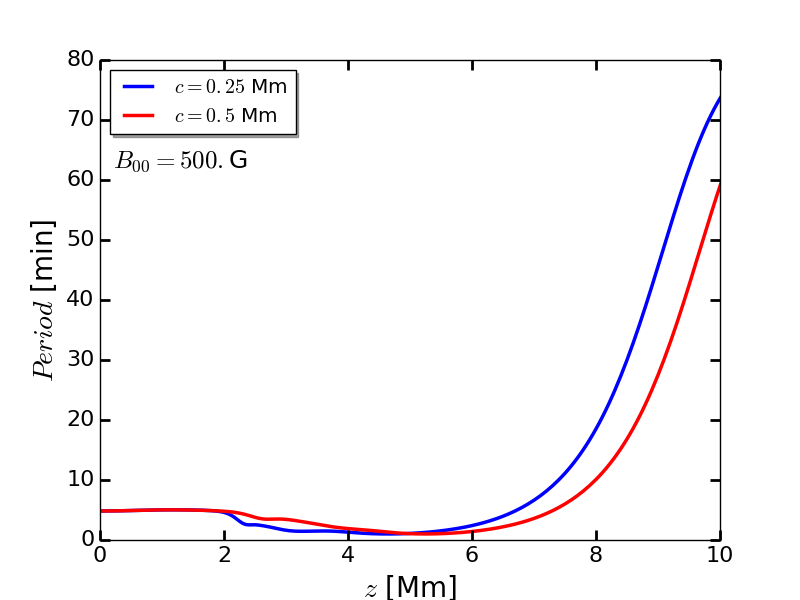}
\caption{Cut-off periods of MAG waves for the two different temperature profiles considering fixed magnetic field intensity $B_{00}$=500.$~$G and a magnetic scale height $l$=4.0$h_\mathrm{b}$,  $h_\mathrm{b}$ being the scale height evaluated at the photospheric base, i.e., $h_\mathrm{b}$=$H(z_0)$ (see Section~\ref{sec:atm-model}).}
\label{fig:cutoffa2}
\end{center}
\end{figure}

Through the wave Eq. (\ref{eq:onda}) we obtain the cut-off periods of MAG waves for a solar stratified atmosphere (Eq. \ref{eq:hseq}). Figure~\ref{fig:cutoffa2} shows  cut-off periods as a function of height for the sharp  (blue line) and smooth (red line) cases, obtained for a magnetic field magnitude at the base of $B_{00}=500~$G and a magnetic scale height of $l=4 h_\mathrm{b}$. The periods at photospheric levels are $P({z_0})=4.8~$min; they become smaller $\sim$[2--$4]~$min at chromospheric and transition region altitudes, and increase up to $P({z_\mathrm{f}})=83.4~$min at coronal levels for both $c$ values. Note that in this domain, the lower the temperature gradient results in a displacement of the larger periods to higher altitude regions in the corona. In other words, for a given atmospheric altitude, cut-off periods are smaller if the transition region slope is shallower.
%

\begin{figure}
\begin{center}
\includegraphics[width=0.55\columnwidth]{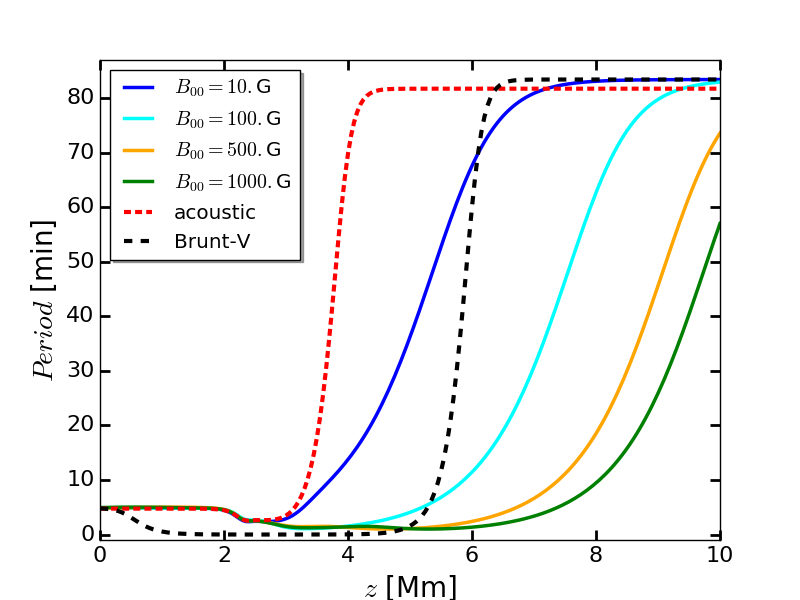}
\caption{Cut-off periods for different magnetic field intensities $B_{00}$, with fixed parameters $c$=0.25$~$Mm and $l$=4.0$h_\mathrm{b}$.}
\label{fig:cutoffB}
\end{center}
\end{figure}

Figure~\ref{fig:cutoffB} allows us to compare the cut-off periods (solid lines) of different magnetic field photospheric intensities. The figure also displays the pure acoustic cut-off period (red dashed line, Eq. \ref{eq:acous}) and the Brunt-V\"ais\"al\"a period (black dashed line). The parameters $c=0.25~$Mm and $l=4.0h_\mathrm{b}$ were used for all cases. The solid blue line corresponds to $B_{00}=10~$G, the black line to $B_{00}=100~$G, the cyan one to $B_{00}=500~$G and the brown one to $1000~$G. The photospheric cut-off periods are shifted towards larger coronal heights while the  magnetic field intensity  increases; moving approximately $2~$Mm per a factor of $10~$G. Below the transition region the periods remain almost constant, $\sim$5$~$min, regardless of the  $B_{00}$ value. For coronal regions (large $z$ values), a larger magnetic field intensity results in shorter cut-off periods. Note also that the period gap between the slow MAG waves and the pure acoustic ones is larger as $B_{00}$ increases. In general, at all heights the slow magneto-acoustic cut-off period is significantly lower than the corresponding acoustic one, except at photospheric heights where they are  almost the same. 
At large coronal heights the pure acoustic and the slow magneto-acoustic periods become similar, while all slow magneto-acoustic periods converge to the Brunt-V\"ais\"al\"a period. Summarising, for transition region and coronal altitudes, intense magnetic field regions, such as sunspots, maintain  small photospheric and chromospheric cut-off period values in comparison with  quiet sun values.
%

\begin{figure}
\begin{center}
\includegraphics[width=0.495\columnwidth]{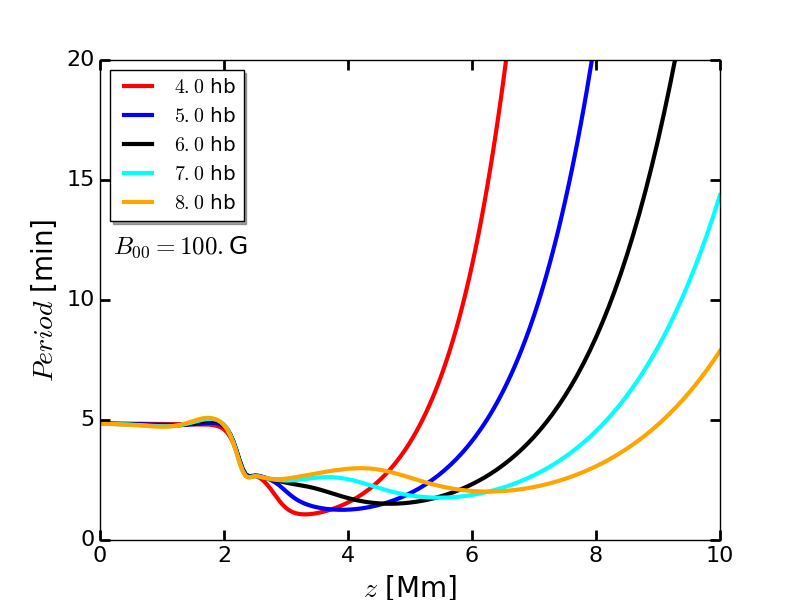}
\includegraphics[width=0.495\columnwidth]{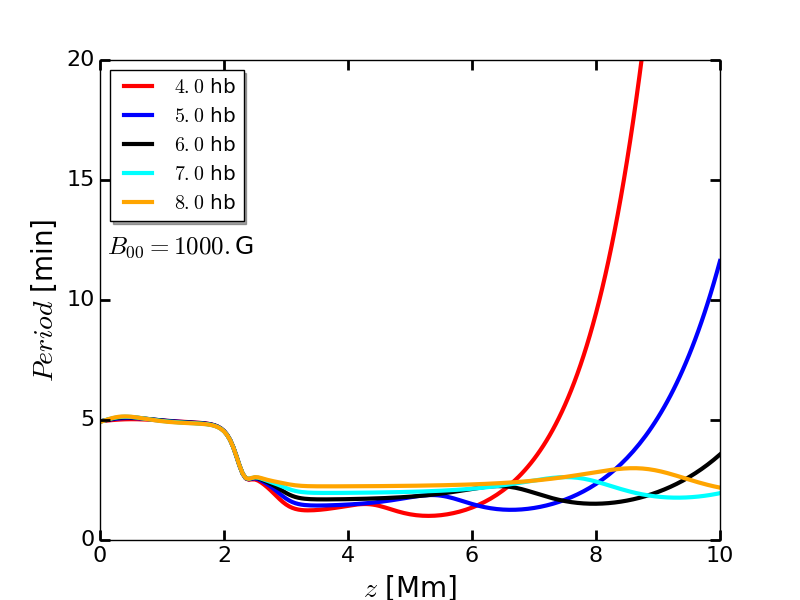}
\caption{Cut-off periods for different magnetic scale heights $l$ (with $l$ in units of the scale height evaluated at the photospheric base $H(z_0) = h_{\mathrm{b}}$), with fixed parameter $c$=0.25$~$Mm and indicated magnetic field intensity $B_{00}$.}
\label{fig:cutoffl}
\end{center}
\end{figure}

We also compare the cut-off periods calculated for different magnetic scale heights $l$ and a given magnetic field intensity value $B_{00}$. Figure~\ref{fig:cutoffl} shows the periods considering $B_{00}=100~$G (top panel) and $1000~$G (bottom panel). We can appreciate from the comparison between panels of the figure, that cut-off periods are shifted towards larger heights for intense magnetic fields, mostly at the transition region and coronal heights. On the other hand, at photospheric heights, the MAG cut-off periods remain almost constant, around $5~$min. 

The colour lines of each panel show the cut-off periods for different tubes, where smaller $l$ values represent more divergent tubes and higher decaying factors of the magnetic field intensity. The cut-off periods are shifted towards higher altitudes for increasing values of $l$, i.e. for magnetic intensities decaying slower with height. Notice that this behaviour is similar to the one found considering different values of $B_{00}$, suggesting that the intensity of the magnetic field along the tube influences the cut-off periods. Figure~\ref{fig:cutoffl} (bottom panel) also suggests that for a thin flux tube with intense magnetic field and a lower degree of divergence (e.g., the umbra of a sunspot), the MAG cut-off period will be almost the same  along the tube that goes from the photosphere up to the corona.
%

\begin{figure}
\begin{center}
\includegraphics[width=0.495\columnwidth]{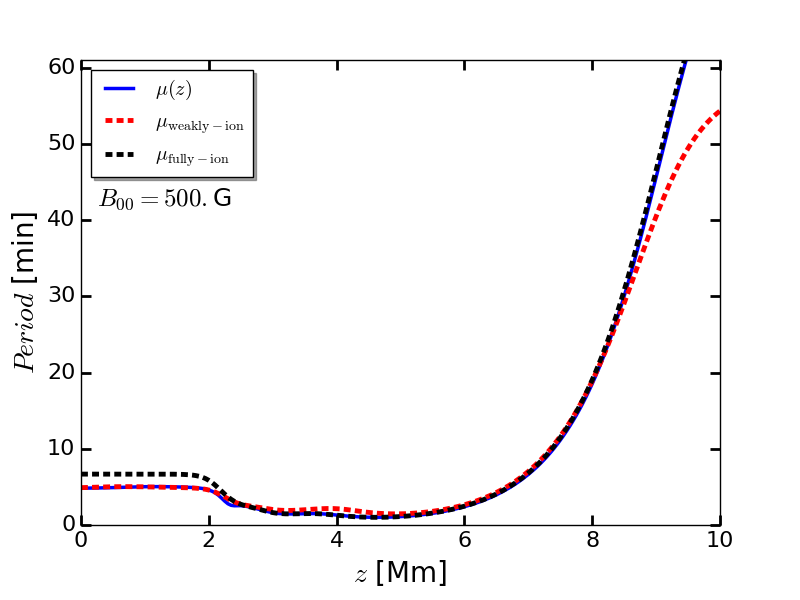}
\includegraphics[width=0.495\columnwidth]{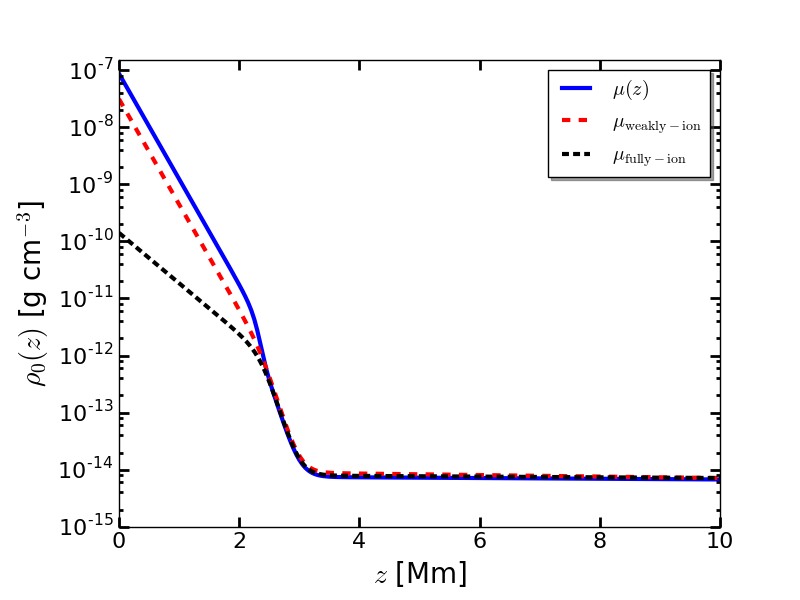}
\caption{a) Cut-off periods comparison using $\mu(z)$, $\mu_\mathrm{w-ion}$=1.29 and $\mu_\mathrm{f-ion}$=0.62. With $l$=4.0$h_\mathrm{b}$, width $c$=0.25$~$Mm and shown magnetic intensity. b) Density profiles calculated with the same parameters of panel (a). The solid blue profile is equivalent to that shown in Figure~\ref{fig:temp_dens-profile}, $\rho_0(z_0)=8.6$$\times$10$^{-8}~$g cm$^{-3}$ and $\rho_0(z_\mathrm{f})=6.7$$\times$10$^{-15}~$g cm$^{-3}$, whereas at the base dashed red is $\rho_0^\mathrm{w-ion}(z_0)=3.1$$\times$10$^{-8}~$g cm$^{-3}$ and dashed black is $\rho_0^\mathrm{f-ion}(z_0)=1.4$$\times$10$^{-10}~$g cm$^{-3}$. Same value for all profiles at $z$=$z_\mathrm{f}$.}
\label{fig:mu}
\end{center}
\end{figure}

Finally, we compare the obtained cut-off periods assuming the mean atomic weight profile $\mu(z)$ given by Eq. (\ref{eq.mu}), with constant $\mu$ profiles, representing a weakly ionised and a fully ionised atmospheres, revisiting \citetalias{costa2018}. The upper panel of Figure~\ref{fig:mu} shows the periods considering the mean atomic weight for $l=4.0\,h_\mathrm{b}$, $c=0.25~$Mm, and a magnetic field intensity of $B_{00}=500~$G. The colour scheme used in the figure is as follows: in solid blue the period values for $\mu(z)$, in dashed red the resulting periods for an overall weakly ionised atmosphere of $\mu_\mathrm{w-ion}$=1.29, and in dashed black the results for an overall fully ionised atmosphere with $\mu_\mathrm{f-ion}$=0.62. The bottom panel shows the  equilibrium density profiles used to calculate the cut-off periods. They all share the same coronal density value at $z_\mathrm{f}$. Note that the consideration of a fully ionised atmosphere leads to the underestimation of the density at photospheric regions by three orders of magnitude. However, the resulting period values are only slightly overestimated, by less than two minutes. At coronal heights the cut-off periods obtained with $\mu(z)$ and $\mu_\mathrm{f-ion}$ coincide. The consideration of an overall weakly ionised atmosphere implies a small underestimation of the photospheric density and an underestimation of periods ($\sim$20$~$min) at coronal levels. Note that the base scale height $h_\mathrm{b}$ is determined by the mean atomic weight at $z$=$z_0$.  However, in order to define the magnetic field scale, $l=4h_\mathrm{b}$, we used the same value for all $\mu$ cases in Figure~\ref{fig:mu}, i.e., the mean atomic weight corresponding to a weakly ionised atmosphere at the base. Otherwise, if we use the corresponding $\mu$ value for each case the overall fully ionised one will have a larger $h_\mathrm{b}$ value and thus a larger $l$, and the cut-off periods represented in dash black will be shifted toward higher altitudes (as it was discussed in Figure~\ref{fig:cutoffl}).
%

\subsection{Comparison with a sunspot observation}

The thin tube approximation is more accurate for large magnetic field intensities, thus we choose a particular sunspot observation to make a comparison with our analytical results. For this task, we chose an AR previously studied for a different date, namely AR 1243  \citep{stekel2014}. An in-depth analysis of the pseudo-periodic phenomena associated to this AR is underway (Sieyra et al., in prep) based on observations acquired on 2011 July 6 in all SDO/AIA channels. Briefly, this AR exhibits a coherent and recurrent arc-shaped intensity disturbance that seems to propagate along several pseudo-open field lines with origin on the sunspot. The intensity disturbances are clearly seen by naked eye in wavelet-processed images \citep{stenborg2013} obtained in several extreme-ultraviolet channels of SDO/AIA. The study carried out by \citet{stekel2014} utilising the majority of the SDO/AIA channels and Sieyra et al. (in prep) suggested that the propagating intensity disturbances observed were the signature of slow-mode MAG waves.

Therefore, to carry out a quick comparison of our theoretical results with actual observations, we just report here on observations from selected channels of the SDO/AIA instrument recorded on 2011 July 3 to highlight the different periodicities that exist at the lowermost layers of the solar atmosphere and the lower corona; namely at 1700~\AA~(continuum ultraviolet light --photosphere and lower chromosphere), 304~\AA~(He~II emission at around 5$\times$10$^4$~K --upper chromosphere and lower transition region) and 171~\AA~(Fe~IX emission at around 6$\times$10$^5$~K --upper transition region and quiet corona).

\begin{figure*}
    \centering
    \includegraphics[width=\textwidth]{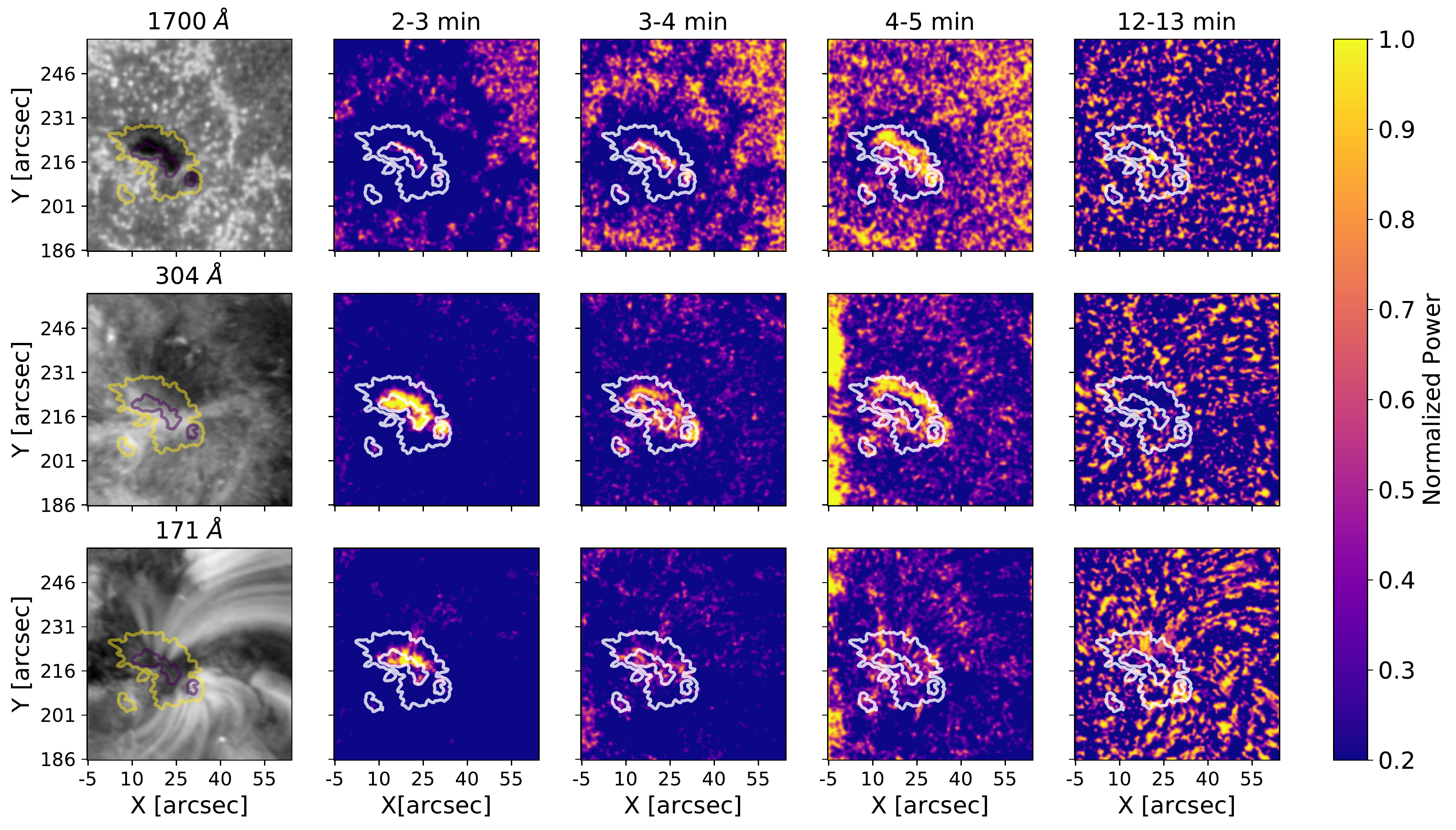}
    \caption{
    Upper row: From left to right, SDO/AIA cropped image at 1700~\AA~(2011 July 3 at 12:00 UT); maps of Lomb-Scargle power in selected periodicity ranges: 2--3 min, 4--5 min, 5--4 min, and 12--13 min, respectively, as computed from the light curves in the 1700$~$\AA~channel. Middle and bottom rows: same as upper row but for observations obtained in the 304$~$\AA~and 171$~$\AA~channels, respectively (extent of the time series: 120 min). The purple and yellow contours correspond to the external umbra and penumbra boundaries taken from  a co-temporal SDO/HMI continuum intensity image (see Figure~\ref{fig:obs_magn}). The white contours on the power maps represent the same as the purple and yellow contours on the power maps.
    }
    \label{fig:obs_30min}
\end{figure*}

\begin{figure}
\begin{center}
\includegraphics[width=0.55\columnwidth]{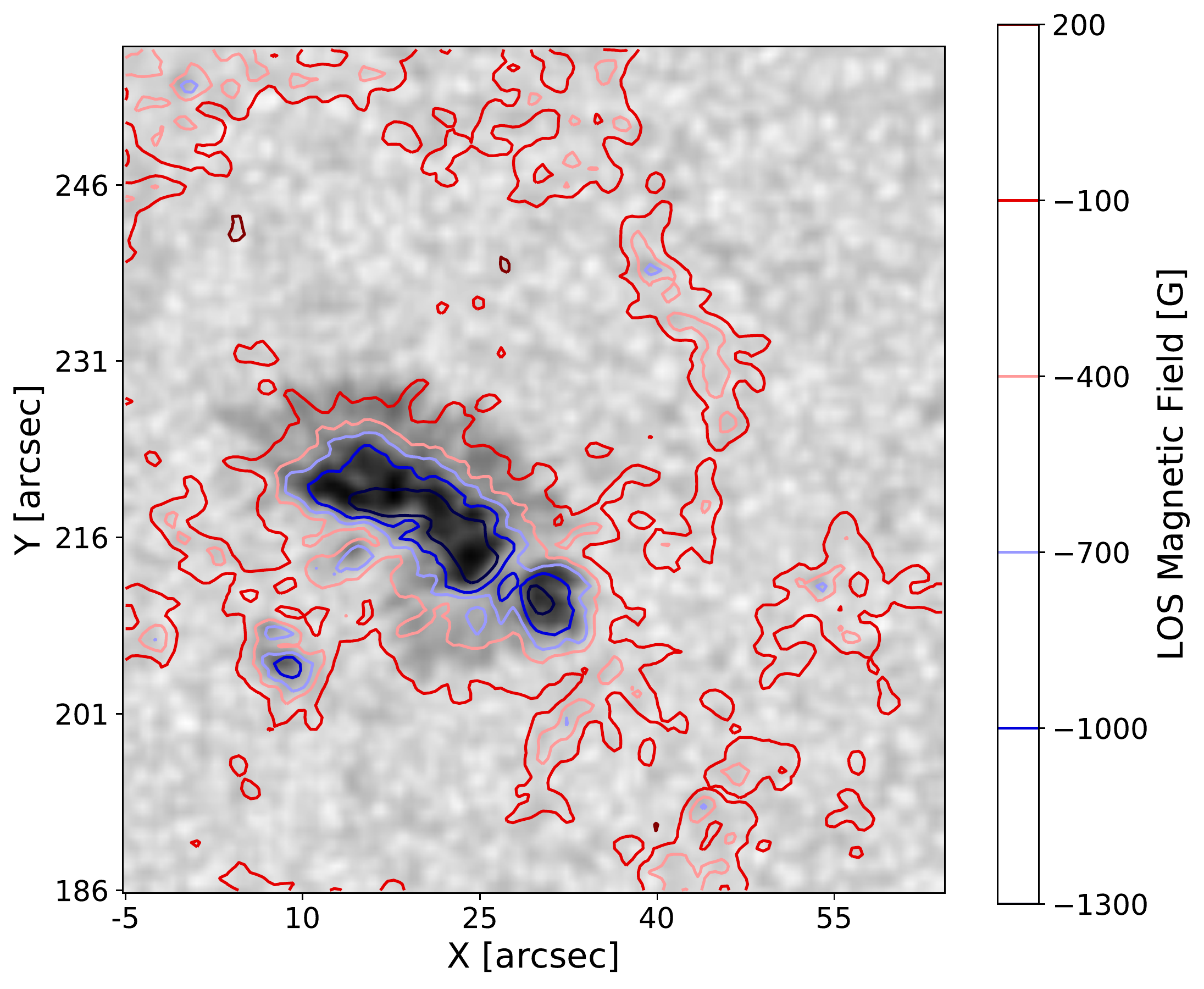}
\caption{Continuum intensity image from SDO/HMI at 12:00 UT and magnetic field contours (in G) taken from LOS magnetogram also from SDO/HMI at 12:00 UT.}
\label{fig:obs_magn}
\end{center}
\end{figure}

In the first column of Figure~\ref{fig:obs_30min} we display three snapshots of the region of interest (ROI) in and around the AR as recorded in the three mentioned SDO/AIA channels. In particular, in the SDO/AIA image of the photosphere at 1700~\AA{} (left-top panel), the dark umbra and grey-dark penumbra of the sunspot and the typical granular pattern from the surrounding photosphere can be seen sharply defined. On the other hand, in the upper transition region--quiet corona counterpart as seen at 171~\AA~(left-bottom panel), the extension of the photospheric magnetic field to coronal heights is evidenced by the ray-like, bright features. The purple and yellow contours in the snapshots delineate the external umbra and penumbra boundaries, which were extracted from a co-temporal continuum image recorded by the Helioseismic and Magnetic Imager \citep[HMI,][]{scherrer2012} on board SDO, with continuum intensity centred at 6173~\AA~(see Figure~\ref{fig:obs_magn}). In Figure~\ref{fig:obs_magn}, the contours indicate the magnetic field strength (as obtained from the line of sight magnetograms from SDO/HMI). Note that the magnetic field intensity in the umbra is $\sim$1000$~$G, then decays up to $\sim$100$~$G in the penumbra and finally takes values between [1--10]$~$G in the quiet sun. The negative sign indicates the magnetic field polarity.

In order to determine and compare the dominant periods present at photospheric, chromospheric and coronal heights in the ROI captured in the images of the left panels of Figure \ref{fig:obs_30min} we considered $120~$min of observations (starting at 12:00 UT). Briefly, the images were treated for analysis using standard routines available in IDL SolarSoft and were corrected to keep the center of the AR at the same position on each channel and all along the time sequence. The continuum images were scaled to the plate scale of 1700~\AA{}, 304~\AA{} y 171~\AA{} images. To look for the dominant periodicities we computed the periodograms from the light curves of the intensity at each pixel of the data sets using the Lomb-Scargle method \citep{press1989}. Since our aim was not to compute the full power spectrum of the time series but to simply characterize the pseudo-periodic nature of the signal observed at relatively high frequencies as permitted by the temporal cadence of the observations with our theoretical model outcome, we restricted the search to periodicities between 1.5 and 15~min \citep[see ][for more details]{ireland2015}.

The panels in columns 2 through 5 on Figure~\ref{fig:obs_30min} display the resulting Lomb-Scargle power (normalized to the maximum power in the [1.5--15]~min range) of the corresponding time series extracted from the SDO/AIA channels and ROI indicated on the first column, in selected 1-min periodicity bins; namely [2--3], [3--4], [4--5], and [12--13]~min, respectively.The white contours delineate the location of the external umbra and penumbra boundaries.

In agreement with well known results \citep[e.g.,][and references therein]{demoortel2012}, Figure~\ref{fig:obs_30min} shows that on averaged the chromospheric and coronal periods (second and third rows) are larger than the photospheric ones (first row), which is also in agreement with the analytical results presented in the Figures  [\ref{fig:cutoffa2}--\ref{fig:cutoffl}]. In the photospheric umbra, where the magnetic field intensity is $\sim$1000$~$G, the dominant periods are between [4--5] min, while in the chromospheric umbra we found mainly lower periods, between [2--3] min. This feature is in accordance with the decrease of the analytical cut-off curve at the beginning of the transition region (see Figure~\ref{fig:cutoffl}). At the coronal level, a concentration of thin flux tubes containing bundles of intense magnetic field lines seemingly anchored to the umbra, exhibit periods between [2--3]$~$min (see bottom row of Figure~\ref{fig:obs_30min}). The flux tubes are initially straight and then bend higher in the corona. In the analytical scenario these regions could be associated with flux tubes characterised by intense magnetic fields and/or less divergent radii, i.e. high $B_{00}$ ($\sim$1000$~$G) and/or $l$ values, where the periods remain low from the photosphere up to the corona (see bottom panel of Figure \ref{fig:cutoffl}). Upwards, when the magnetic flux tubes become more divergent and/or more inclined, larger periods are measured along the features that trace these magnetic structures. Towards the penumbra, around the umbra sunspot, where the flux tubes have less intense magnetic fields and/or are more divergent, the periods increase (see upper panel of Figure \ref{fig:cutoffl}), in agreement with others observational studies \citep{madsen2015,sych2014,freij2014,jess2013}. Within the analytical framework this fact could be attributed to small $l$ values and/or $B_{00}$, as can be seen in Figure \ref{fig:obs_magn}.

Summarising, in the photospheric, chromospheric and coronal umbra, where the magnetic flux tubes are expected to be thin, straight and with strong magnetic field intensity as was modelled by large $B_{00}$ and $l$ values, the observational and analytical results coincide displaying low periods ($\leq$5$~$min). At the  penumbra, where the magnetic flux tubes are expected to have a weaker magnetic field and/or be more divergent, which may be considered to have a smaller intensity $B_{00}$ and/or $l$ values, the observed periods are higher than the corresponding umbral ones.

\section{Conclusions}
In this work, we derived  the cut-off periods of slow MAG waves in the solar atmosphere, from the photosphere through the lower corona, using a thin flux tube model which includes a temperature and mean atomic weight stratification with height. In particular, we studied the effect of a strong temperature gradient  on the value of the cut-off periods \citep{2015ARep...59..959D}, the variations of the mean atomic mass, as well as the strength and degree of divergence of the magnetic thin tubes that emerge from the photosphere at localised magnetic features, such as sunspots, and that expand towards the corona. The periods obtained are in accordance with known results from  literature \citep[e.g. \citetalias{costa2018},][]{afanasyev2015,sych2014,demoortel2012,2006roberts}. We obtained results that show that the cut-off values vary with height and mostly depend on the atmospheric stratification and intensity of the magnetic field.

As the accuracy of the model is mostly confined to large magnetic field values, we compared our results with observational data obtained from a sunspot and its coronal counterpart. For these type of intense magnetic field regions, we showed that magnetic tube regions almost straight (i.e., regions with larger values of $l$) are associated with low periods ($\sim$5$~$min). Also, as the  magnetic field strength remains sufficiently high, the  magnetic tubes remain almost straight and the periods remain relatively small at higher altitudes in the solar atmosphere, e.g., small  periods are characteristics at all solar atmospheric levels for structures above umbra sunspots \citep[e.g.][]{marsh2006,demoortel2002}. Whereas more divergent flux tubes or less intense magnetic fields (typically found at high coronal levels) are associated with longer periods.

Sunspot images are complex objects, where superposition of different atmospheric layers and frequencies together with projection issues occur, nonetheless this analytic cut-off frequency scenario can capture the main sunspot period features.

\section*{Acknowledgements}

E.Z. acknowledges support from Funda\c{c}\~ao de Amparo \`a Pesquisa do Estado de S\~ao Paulo (FAPESP) through the post-doc grant 2018/25177-4 (Brasil). Part of this work was done with the support from a post-doc grant of CONICET-IATE (Argentina). M.V.S. acknowledges support from CONICET (Argentina).
A.E acknowledges support from CONACYT (Mexico) grant 167611, DGAPA-PAPIIT (UNAM) grant IG-RG 100516, and support from the DGAPA-PASPA (UNAM) program. G.S. acknowledges the support of NASA (grant NNG17PP271 and contract NNG09EK11I to NRL). {\sc Chianti} is a collaborative project involving George Mason University, the University of Michigan (USA), University of Cambridge (UK) and NASA Goddard Space Flight Center (USA).

\appendix
\section{} \label{s:appendix}

With the aim of obtaining an expression for the cut-off period, the wave Eq. (\ref{eq:onda}) is reduced to a Klein-Gordon form \citep[see, e.g.,][]{1953mtp..book.....M}. The main idea of this procedure is to eliminate the first partial derivative of the longitudinal velocity ($\partial u/\partial z$) in the wave Eq. (\ref{eq:onda}), i.e., the term accompanied by the function $K_1(z)$. Thus, a change of variable is made, proposing $u(z,t)=e^{\psi (z)}U(z,t)$. After replacing it in the wave equation and identifying $\psi'=K_1/(2c_{\mathrm{T}}^2)$, a Klein-Gordon equation for the new variable $U(z,t)$ is obtained
\begin{eqnarray} 
\begin{aligned}
\frac{\partial^2 U}{\partial t^2}-c_{\mathrm{T}}^2 \frac{\partial^2 U}{\partial z^2}+\omega^2_u U=0, 
\end{aligned}
\end{eqnarray}
which contain explicitly the cut-off frequency that is equal to
\begin{eqnarray} 
\begin{aligned}
\omega^2_u(z) = -c_{\mathrm{T}}^2 \left( \psi' \right)^2 -c_{\mathrm{T}}^2 \psi'' + K_1 \psi' + K_2. 
\end{aligned}
\end{eqnarray}

The general expression for the cut-off frequency squared $\omega^2_u(z)$ is:
\begin{eqnarray} \label{eq:cutoff_gen}
\begin{aligned}
\omega_u^2(z) = \ & c_{\mathrm{T}}^2 \left(1-\frac{1}{2} \frac{V_{\mathrm{A}}^2-c_0^2}{V_{\mathrm{A}}^2+c_0^2}\right) \frac{B_0''}{B_0} \\ &
- \frac{c_{\mathrm{T}}^2}{2} \frac{V_{\mathrm{A}}^2-c_0^2}{V_{\mathrm{A}}^2+c_0^2} \left(1-\frac{1}{2} \frac{V_{\mathrm{A}}^2-c_0^2}{V_{\mathrm{A}}^2+c_0^2}\right) \left( \frac{B_0'}{B_0} \right)^2  \\ &
+ g\frac{c_{\mathrm{T}}^4}{c_0^4} \left[3\frac{c_0^2}{V_{\mathrm{A}}^2}+1-\gamma + \frac{\gamma}{2}\frac{V_{\mathrm{A}}^2-c_0^2}{V_{\mathrm{A}}^2+c_0^2} \right. \\ &
\left. + \frac{2}{g}\frac{c_0^2 c_{\mathrm{T}}^2}{V_{\mathrm{A}}^2} \left(\frac{c_0'}{c_0} - \frac{V_{\mathrm{A}}'}{V_{\mathrm{A}}}\right)\right] \frac{B_0'}{B_0} - g\gamma\frac{c_{\mathrm{T}}^4}{c_0^4} \left(1-\frac{c_{\mathrm{T}}^2}{c_0^2}\right) \frac{V_{\mathrm{A}}'}{V_{\mathrm{A}}} \\ &
+ g \frac{c_{\mathrm{T}}^2}{c_0^2} \left( \gamma\frac{c_{\mathrm{T}}^2}{c_0^2} + 2\frac{c_0^2}{V_{\mathrm{A}}^2} + \gamma\frac{c_{\mathrm{T}}^4}{V_{\mathrm{A}}^2 c_0^2} \right) \frac{c_0'}{c_0} \\ &
+ g^2\frac{c_{\mathrm{T}}^4}{c_0^2 V_{\mathrm{A}}^2} \left( \frac{\gamma-1}{c_{\mathrm{T}}^2}+\frac{\gamma}{c_0^2}+\frac{\gamma^2}{4} \frac{V_{\mathrm{A}}^2 c_{\mathrm{T}}^2}{c_0^6} \right).
\end{aligned}
\end{eqnarray}
The cut-off frequency was written in terms of generals magnetic field $B_0(z)$, Alv\'en speed $V_{\mathrm{A}}(z)$, sound speed $c_0(z)$, their derivatives and tube speed $c_{\mathrm{T}}(z)$. In its turn the explicit expression for the cut-off period is obtained from $P(z)=2\pi/\omega_u(z)$.

The Eq. (\ref{eq:cutoff_gen}) is valid for the $z$-dependent equation of state (\ref{eq:gideal}), arbitrary although consistent equilibrium profiles for density $\rho_0(z)$, temperature $T_0(z)$ and mean atomic weight $\mu(z)$, and a general magnetic field $B_0(z)$ (valid in the framework of the thin flux tube model).



\bibliographystyle{mnras}
\bibliography{references} 

\begin{thebibliography}{}
\makeatletter
\relax
\def\mn@urlcharsother{\let\do\@makeother \do\$\do\&\do\#\do\^\do\_\do\%\do\~}
\def\mn@doi{\begingroup\mn@urlcharsother \@ifnextchar [ {\mn@doi@}
  {\mn@doi@[]}}
\def\mn@doi@[#1]#2{\def\@tempa{#1}\ifx\@tempa\@empty \href
  {http://dx.doi.org/#2} {doi:#2}\else \href {http://dx.doi.org/#2} {#1}\fi
  \endgroup}
\def\mn@eprint#1#2{\mn@eprint@#1:#2::\@nil}
\def\mn@eprint@arXiv#1{\href {http://arxiv.org/abs/#1} {{\tt arXiv:#1}}}
\def\mn@eprint@dblp#1{\href {http://dblp.uni-trier.de/rec/bibtex/#1.xml}
  {dblp:#1}}
\def\mn@eprint@#1:#2:#3:#4\@nil{\def\@tempa {#1}\def\@tempb {#2}\def\@tempc
  {#3}\ifx \@tempc \@empty \let \@tempc \@tempb \let \@tempb \@tempa \fi \ifx
  \@tempb \@empty \def\@tempb {arXiv}\fi \@ifundefined
  {mn@eprint@\@tempb}{\@tempb:\@tempc}{\expandafter \expandafter \csname
  mn@eprint@\@tempb\endcsname \expandafter{\@tempc}}}

\bibitem[\protect\citeauthoryear{{Afanasyev} \& {Nakariakov}}{{Afanasyev} \&
  {Nakariakov}}{2015}]{afanasyev2015}
{Afanasyev} A.~N.,  {Nakariakov} V.~M.,  2015, \mn@doi [\aap]
  {10.1051/0004-6361/201526530}, \href
  {http://adsabs.harvard.edu/abs/2015A%26A...582A..57A} {582, A57}

\bibitem[\protect\citeauthoryear{{Bogdan} \& {Judge}}{{Bogdan} \&
  {Judge}}{2006}]{2006RSPTA.364..313B}
{Bogdan} T.~J.,  {Judge} P.~G.,  2006, \mn@doi [Philosophical Transactions of
  the Royal Society of London Series A] {10.1098/rsta.2005.1701}, \href
  {https://ui.adsabs.harvard.edu/abs/2006RSPTA.364..313B} {364, 313}

\bibitem[\protect\citeauthoryear{{Costa}, {Schneiter}  \& {Zurbriggen}}{{Costa}
  et~al.}{2018}]{costa2018}
{Costa} A.,  {Schneiter} M.,   {Zurbriggen} E.,  2018, \mn@doi [\mnras]
  {10.1093/mnras/sty1828}, \href
  {http://adsabs.harvard.edu/abs/2018MNRAS.480..623C} {480, 623}

\bibitem[\protect\citeauthoryear{{De Moortel} \& {Nakariakov}}{{De Moortel} \&
  {Nakariakov}}{2012}]{demoortel2012}
{De Moortel} I.,  {Nakariakov} V.~M.,  2012, \mn@doi [Philosophical
  Transactions of the Royal Society of London Series A]
  {10.1098/rsta.2011.0640}, \href
  {http://adsabs.harvard.edu/abs/2012RSPTA.370.3193D} {370, 3193}

\bibitem[\protect\citeauthoryear{{De Moortel}, {Ireland}, {Hood}  \&
  {Walsh}}{{De Moortel} et~al.}{2002}]{demoortel2002}
{De Moortel} I.,  {Ireland} J.,  {Hood} A.~W.,   {Walsh} R.~W.,  2002, \mn@doi
  [\aap] {10.1051/0004-6361:20020436}, \href
  {http://adsabs.harvard.edu/abs/200226A...387L..13D} {387, L13}

\bibitem[\protect\citeauthoryear{{Dere}, {Landi}, {Mason}, {Monsignori Fossi}
  \& {Young}}{{Dere} et~al.}{1997}]{1997A&AS..125..149D}
{Dere} K.~P.,  {Landi} E.,  {Mason} H.~E.,  {Monsignori Fossi} B.~C.,   {Young}
  P.~R.,  1997, \mn@doi [\aaps] {10.1051/aas:1997368}, \href
  {https://ui.adsabs.harvard.edu/abs/1997A%26AS..125..149D} {125, 149}

\bibitem[\protect\citeauthoryear{{Dere}, {Del Zanna}, {Young}, {Landi}  \&
  {Sutherland}}{{Dere} et~al.}{2019}]{2019ApJS..241...22D}
{Dere} K.~P.,  {Del Zanna} G.,  {Young} P.~R.,  {Landi} E.,   {Sutherland}
  R.~S.,  2019, \mn@doi [\apjs] {10.3847/1538-4365/ab05cf}, \href
  {https://ui.adsabs.harvard.edu/abs/2019ApJS..241...22D} {241, 22}

\bibitem[\protect\citeauthoryear{{Deres} \& {Anfinogentov}}{{Deres} \&
  {Anfinogentov}}{2015}]{2015ARep...59..959D}
{Deres} A.~S.,  {Anfinogentov} S.~A.,  2015, \mn@doi [Astronomy Reports]
  {10.1134/S1063772915100017}, \href
  {https://ui.adsabs.harvard.edu/abs/2015ARep...59..959D} {59, 959}

\bibitem[\protect\citeauthoryear{{Fontenla}, {Curdt}, {Haberreiter}, {Harder}
  \& {Tian}}{{Fontenla} et~al.}{2009}]{2009ApJ...707..482F}
{Fontenla} J.~M.,  {Curdt} W.,  {Haberreiter} M.,  {Harder} J.,   {Tian} H.,
  2009, \mn@doi [\apj] {10.1088/0004-637X/707/1/482}, \href
  {https://ui.adsabs.harvard.edu/abs/2009ApJ...707..482F} {707, 482}

\bibitem[\protect\citeauthoryear{{Freij}, {Scullion}, {Nelson}, {Mumford},
  {Wedemeyer}  \& {Erd{\'e}lyi}}{{Freij} et~al.}{2014}]{freij2014}
{Freij} N.,  {Scullion} E.~M.,  {Nelson} C.~J.,  {Mumford} S.,  {Wedemeyer} S.,
    {Erd{\'e}lyi} R.,  2014, \mn@doi [\apj] {10.1088/0004-637X/791/1/61}, \href
  {http://adsabs.harvard.edu/abs/2014ApJ...791...61F} {791, 61}

\bibitem[\protect\citeauthoryear{{Ireland}, {McAteer}  \& {Inglis}}{{Ireland}
  et~al.}{2015}]{ireland2015}
{Ireland} J.,  {McAteer} R.~T.~J.,   {Inglis} A.~R.,  2015, \mn@doi [\apj]
  {10.1088/0004-637X/798/1/1}, \href
  {https://ui.adsabs.harvard.edu/abs/2015ApJ...798....1I} {798, 1}

\bibitem[\protect\citeauthoryear{{Jess}, {De Moortel}, {Mathioudakis},
  {Christian}, {Reardon}, {Keys}  \& {Keenan}}{{Jess} et~al.}{2012}]{jess2012}
{Jess} D.~B.,  {De Moortel} I.,  {Mathioudakis} M.,  {Christian} D.~J.,
  {Reardon} K.~P.,  {Keys} P.~H.,   {Keenan} F.~P.,  2012, \mn@doi [\apj]
  {10.1088/0004-637X/757/2/160}, \href
  {http://adsabs.harvard.edu/abs/2012ApJ...757..160J} {757, 160}

\bibitem[\protect\citeauthoryear{{Jess}, {Reznikova}, {Van Doorsselaere},
  {Keys}  \& {Mackay}}{{Jess} et~al.}{2013}]{jess2013}
{Jess} D.~B.,  {Reznikova} V.~E.,  {Van Doorsselaere} T.,  {Keys} P.~H.,
  {Mackay} D.~H.,  2013, \mn@doi [\apj] {10.1088/0004-637X/779/2/168}, \href
  {http://adsabs.harvard.edu/abs/2013ApJ...779..168J} {779, 168}

\bibitem[\protect\citeauthoryear{{Khomenko}}{{Khomenko}}{2017}]{2017PPCF...59a4038K}
{Khomenko} E.,  2017, \mn@doi [Plasma Physics and Controlled Fusion]
  {10.1088/0741-3335/59/1/014038}, \href
  {https://ui.adsabs.harvard.edu/abs/2017PPCF...59a4038K} {59, 014038}

\bibitem[\protect\citeauthoryear{{Khomenko} \& {Collados}}{{Khomenko} \&
  {Collados}}{2015}]{2015LRSP...12....6K}
{Khomenko} E.,  {Collados} M.,  2015, \mn@doi [Living Reviews in Solar Physics]
  {10.1007/lrsp-2015-6}, \href
  {https://ui.adsabs.harvard.edu/abs/2015LRSP...12....6K} {12, 6}

\bibitem[\protect\citeauthoryear{{King}, {Nakariakov}, {Deluca}, {Golub}  \&
  {McClements}}{{King} et~al.}{2003}]{2003A&A...404L...1K}
{King} D.~B.,  {Nakariakov} V.~M.,  {Deluca} E.~E.,  {Golub} L.,   {McClements}
  K.~G.,  2003, \mn@doi [\aap] {10.1051/0004-6361:20030763}, \href
  {https://ui.adsabs.harvard.edu/abs/2003A&A...404L...1K} {404, L1}

\bibitem[\protect\citeauthoryear{{Lamb}}{{Lamb}}{1932}]{1932hydr.book.....L}
{Lamb} H.,  1932, {Hydrodynamics}.
Dover, New York.

\bibitem[\protect\citeauthoryear{{Lemen} et~al.,}{{Lemen}
  et~al.}{2012}]{lemen2012}
{Lemen} J.~R.,  et~al., 2012, \mn@doi [\solphys] {10.1007/s11207-011-9776-8},
  \href {http://adsabs.harvard.edu/abs/2012SoPh..275...17L} {275, 17}

\bibitem[\protect\citeauthoryear{{Madsen}, {Tian}  \& {DeLuca}}{{Madsen}
  et~al.}{2015}]{madsen2015}
{Madsen} C.~A.,  {Tian} H.,   {DeLuca} E.~E.,  2015, \mn@doi [\apj]
  {10.1088/0004-637X/800/2/129}, \href
  {http://adsabs.harvard.edu/abs/2015ApJ...800..129M} {800, 129}

\bibitem[\protect\citeauthoryear{{Marsh} \& {Walsh}}{{Marsh} \&
  {Walsh}}{2006}]{marsh2006}
{Marsh} M.~S.,  {Walsh} R.~W.,  2006, \mn@doi [\apj] {10.1086/501450}, \href
  {http://adsabs.harvard.edu/abs/2006ApJ...643..540M} {643, 540}

\bibitem[\protect\citeauthoryear{{Morse} \& {Feshbach}}{{Morse} \&
  {Feshbach}}{1953}]{1953mtp..book.....M}
{Morse} P.~M.,  {Feshbach} H.,  1953, {Methods of theoretical physics Part I}.
Mc Graw Hill, New York.

\bibitem[\protect\citeauthoryear{{Nakariakov}}{{Nakariakov}}{2006}]{2006RSPTA.364..473N}
{Nakariakov} V.~M.,  2006, \mn@doi [Philosophical Transactions of the Royal
  Society of London Series A] {10.1098/rsta.2005.1711}, \href
  {https://ui.adsabs.harvard.edu/abs/2006RSPTA.364..473N} {364, 473}

\bibitem[\protect\citeauthoryear{{Pesnell}, {Thompson}  \&
  {Chamberlin}}{{Pesnell} et~al.}{2012}]{pesnell2012}
{Pesnell} W.~D.,  {Thompson} B.~J.,   {Chamberlin} P.~C.,  2012, \mn@doi [Solar
  Physics] {10.1007/s11207-011-9841-3}, 275, 3

\bibitem[\protect\citeauthoryear{{Press} \& {Rybicki}}{{Press} \&
  {Rybicki}}{1989}]{press1989}
{Press} W.~H.,  {Rybicki} G.~B.,  1989, \mn@doi [\apj] {10.1086/167197}, \href
  {https://ui.adsabs.harvard.edu/abs/1989ApJ...338..277P} {338, 277}

\bibitem[\protect\citeauthoryear{{Reznikova}, {Shibasaki}, {Sych}  \&
  {Nakariakov}}{{Reznikova} et~al.}{2012}]{reznikova2012}
{Reznikova} V.~E.,  {Shibasaki} K.,  {Sych} R.~A.,   {Nakariakov} V.~M.,  2012,
  \mn@doi [\apj] {10.1088/0004-637X/746/2/119}, \href
  {http://adsabs.harvard.edu/abs/2012ApJ...746..119R} {746, 119}

\bibitem[\protect\citeauthoryear{{Roberts}}{{Roberts}}{2006}]{2006roberts}
{Roberts} B.,  2006, \mn@doi [Philosophical Transactions of the Royal Society
  of London Series A] {10.1098/rsta.2005.1709}, \href
  {http://adsabs.harvard.edu/abs/2006RSPTA.364..447R} {364, 447}

\bibitem[\protect\citeauthoryear{{Roberts} \& {Ulmschneider}}{{Roberts} \&
  {Ulmschneider}}{1997}]{1997LNP...489...75R}
{Roberts} B.,  {Ulmschneider} P.,  1997, {Dynamics of Flux Tubes in the Solar
  Atmosphere: Theory}.
Proceedings of the 8th European Meeting on Solar Physics Held at Halkidiki.
  Springer-Verlag Berlin Heidelberg New York. Also Lecture Notes in Physics,
  volume 489, p.75, p.~75, \mn@doi{10.1007/BFb0105671}

\bibitem[\protect\citeauthoryear{{Roberts} \& {Webb}}{{Roberts} \&
  {Webb}}{1978}]{1978SoPh...56....5R}
{Roberts} B.,  {Webb} A.~R.,  1978, \mn@doi [\solphys] {10.1007/BF00152630},
  \href {https://ui.adsabs.harvard.edu/abs/1978SoPh...56....5R} {56, 5}

\bibitem[\protect\citeauthoryear{{Roberts} \& {Webb}}{{Roberts} \&
  {Webb}}{1979}]{1979SoPh...64...77R}
{Roberts} B.,  {Webb} A.~R.,  1979, \mn@doi [\solphys] {10.1007/BF00151117},
  \href {https://ui.adsabs.harvard.edu/abs/1979SoPh...64...77R} {64, 77}

\bibitem[\protect\citeauthoryear{{Sakurai}, {Ichimoto}, {Raju}  \&
  {Singh}}{{Sakurai} et~al.}{2002}]{2002SoPh..209..265S}
{Sakurai} T.,  {Ichimoto} K.,  {Raju} K.~P.,   {Singh} J.,  2002, \mn@doi
  [\solphys] {10.1023/A:1021297313448}, \href
  {https://ui.adsabs.harvard.edu/abs/2002SoPh..209..265S} {209, 265}

\bibitem[\protect\citeauthoryear{{Scherrer} et~al.,}{{Scherrer}
  et~al.}{2012}]{scherrer2012}
{Scherrer} P.~H.,  et~al., 2012, \mn@doi [\solphys]
  {10.1007/s11207-011-9834-2}, \href
  {https://ui.adsabs.harvard.edu/abs/2012SoPh..275..207S} {275, 207}

\bibitem[\protect\citeauthoryear{{Solanki}, {Inhester}  \&
  {Sch{\"u}ssler}}{{Solanki} et~al.}{2006}]{solanki2006}
{Solanki} S.~K.,  {Inhester} B.,   {Sch{\"u}ssler} M.,  2006, \mn@doi [Reports
  on Progress in Physics] {10.1088/0034-4885/69/3/R02}, \href
  {http://adsabs.harvard.edu/abs/2006RPPh...69..563S} {69, 563}

\bibitem[\protect\citeauthoryear{{Soler}, {Terradas}, {Oliver}  \&
  {Ballester}}{{Soler} et~al.}{2017}]{2017ApJ...840...20S}
{Soler} R.,  {Terradas} J.,  {Oliver} R.,   {Ballester} J.~L.,  2017, \mn@doi
  [\apj] {10.3847/1538-4357/aa6d7f}, \href
  {https://ui.adsabs.harvard.edu/abs/2017ApJ...840...20S} {840, 20}

\bibitem[\protect\citeauthoryear{{Srivastava} \& {Dwivedi}}{{Srivastava} \&
  {Dwivedi}}{2010}]{2010NewA...15....8S}
{Srivastava} A.~K.,  {Dwivedi} B.~N.,  2010, \mn@doi [\na]
  {10.1016/j.newast.2009.05.006}, \href
  {https://ui.adsabs.harvard.edu/abs/2010NewA...15....8S} {15, 8}

\bibitem[\protect\citeauthoryear{{Stekel}, {Stenborg}  \& {Dal Lago}}{{Stekel}
  et~al.}{2014}]{stekel2014}
{Stekel} T.~R.~C.,  {Stenborg} G.,   {Dal Lago} A.,  2014, in AGU Fall Meeting
  Abstracts. pp SH13B--4110

\bibitem[\protect\citeauthoryear{{Stenborg}, {Stekel}, {Vourlidas}  \&
  {Howard}}{{Stenborg} et~al.}{2013}]{stenborg2013}
{Stenborg} G.,  {Stekel} T.,  {Vourlidas} A.,   {Howard} R.,  2013, in SDO-3:
  Exploring the Network of SDO Science. p.~55

\bibitem[\protect\citeauthoryear{{Sych} \& {Nakariakov}}{{Sych} \&
  {Nakariakov}}{2014}]{sych2014}
{Sych} R.,  {Nakariakov} V.~M.,  2014, \mn@doi [\aap]
  {10.1051/0004-6361/201424049}, \href
  {https://ui.adsabs.harvard.edu/abs/2014A&A...569A..72S} {569, A72}

\bibitem[\protect\citeauthoryear{{Thomas}}{{Thomas}}{1982}]{thomas1982}
{Thomas} J.~H.,  1982, \mn@doi [\apj] {10.1086/160471}, \href
  {http://adsabs.harvard.edu/abs/1982ApJ...262..760T} {262, 760}

\bibitem[\protect\citeauthoryear{{Vernazza}, {Avrett}  \& {Loeser}}{{Vernazza}
  et~al.}{1981}]{vernazza1981}
{Vernazza} J.~E.,  {Avrett} E.~H.,   {Loeser} R.,  1981, \mn@doi [\apjs]
  {10.1086/190731}, \href {http://adsabs.harvard.edu/abs/1981ApJS...45..635V}
  {45, 635}

\bibitem[\protect\citeauthoryear{{Yuan}, {Sych}, {Reznikova}  \&
  {Nakariakov}}{{Yuan} et~al.}{2014}]{yuan2014}
{Yuan} D.,  {Sych} R.,  {Reznikova} V.~E.,   {Nakariakov} V.~M.,  2014, \mn@doi
  [\aap] {10.1051/0004-6361/201220208}, \href
  {http://adsabs.harvard.edu/abs/2014A%26A...561A..19Y} {561, A19}

\bibitem[\protect\citeauthoryear{{Zavershinskii}, {Kolotkov}, {Nakariakov},
  {Molevich}  \& {Ryashchikov}}{{Zavershinskii}
  et~al.}{2019}]{2019PhPl...26h2113Z}
{Zavershinskii} D.~I.,  {Kolotkov} D.~Y.,  {Nakariakov} V.~M.,  {Molevich}
  N.~E.,   {Ryashchikov} D.~S.,  2019, \mn@doi [Physics of Plasmas]
  {10.1063/1.5115224}, \href
  {https://ui.adsabs.harvard.edu/abs/2019PhPl...26h2113Z} {26, 082113}

\bibitem[\protect\citeauthoryear{{Zhugzhda}}{{Zhugzhda}}{1996}]{zhugzhda1996}
{Zhugzhda} Y.~D.,  1996, \mn@doi [Physics of Plasmas] {10.1063/1.871836}, \href
  {http://adsabs.harvard.edu/abs/1996PhPl....3...10Z} {3, 10}

\bibitem[\protect\citeauthoryear{{Zhugzhda} \& {Dzhalilov}}{{Zhugzhda} \&
  {Dzhalilov}}{1984}]{zhugzhda1984}
{Zhugzhda} I.~D.,  {Dzhalilov} N.~S.,  1984, \aap, \href
  {http://adsabs.harvard.edu/abs/1984A%26A...132...45Z} {132, 45}

\makeatother
\end{thebibliography}








\bsp	
\label{lastpage}
\end{document}